\documentclass[fleqn,11pt]{article}
\usepackage{amsmath}
\usepackage{latexsym}
\usepackage{rotate,epsfig}
\usepackage{lscape}
\numberwithin{equation}{section}

\newtheorem{theorem}{\bf Theorem}[section]
\newtheorem{proposition}[theorem]{\bf Proposition}

\newtheorem{corollary}[theorem]{\bf Corollary}
    {\par \noindent {\bf Remark }}%
    {\par \indent}
    {\par \noindent {\bf Proof}}%
    {\par \indent}
\renewcommand{\vartheta}{\tau}
\renewcommand{\log}{\ln}
\begin{document}
\title{{\bf Explicit expressions for European option pricing under a generalized skew normal distribution}}
\author{{\bf Mahdi Doostparast}\footnote{
\text{E-mail address:} {\it doostparast@math.um.ac.ir}}
   \\
{\small {\it  Department of Statistics, School of
Mathematical Sciences,}}\vspace{-0.2cm}\\ {\small {\it Ferdowsi University of Mashhad, P. O. Box 91775-1159,  Mashhad,\ Iran }}\\
}
\date{}
\maketitle
\begin{abstract}
Under a generalized skew normal distribution, we consider the problem of European option pricing. Existence of the martingale measure is proved. An explicit expression for a given European option price is presented in terms of the cumulative distribution function of the univariate skew normal and the bivariate standard normal distributions. Some special cases are investigated in a greater detail. To carry out the sensitivity of the option price to the skew parameters, numerical methods are applied. Some concluding remarks and further works are given. The results obtained are extension of the results provided by \cite{CS}.
\end{abstract}
\section{Introduction}\label{intro}
A call (put) option
is the right to buy (sell) a particular asset for a strike price at a specified time in the future. There are various types of option. The most common one is European options (EOs) which can only be exercised on the maturity date.  \cite{BlackScholes1973} assumed a geometric Brownian motion for the underlying asset and derived a closed form for fair price of a given European option (EO), known as {\em Black-Scholes option pricing formula}. It is one of the major successes of modern financial economics. But empirical evidences showed that there are systematic pricing errors when compared to observed option prices. For example, \cite{CorradoSu1997} present evidence of systematic mispricing of the Black-Scholes model when the log-returns of the underlying asset are skewed and leptokurtic, typically underpricing options that are deep in-the-money and overpricing options that are out-of-money. To resolve the mispricing problem, Black-Scholes option price has been extended along with several directions. For example,
\cite{CS} assumed that the underlying stock price
process $\{S(t),t\geq 0\}$ follows a geometric Azzalini skew
Brownian motion. More precisely, let
\begin{equation}\label{model:CS}
S(t)=S(0)\exp\left\{\mu
t+\sigma\sqrt{t}Z_\lambda\right\},
\end{equation}
where the random variable
$Z_\lambda$ has a skew normal distribution, denoted by
$SN(\lambda)$, with probability density function (pdf)
\[\phi(x;\lambda)=2\phi(x)\Phi(\lambda x),\ \ \
-\infty<x<+\infty,\] 
where $\phi(x)$ and $\Phi(x)$ are the pdf
and the cumulative distribution function (cdf) of the standard
normal distribution, respectively, i.e.
$$\phi(x)=\frac{1}{2\pi}\exp\left\{-\frac{x^2}{2}\right\},\ \ \ \mbox{and}\ \ \ \Phi(x)=\int_{-\infty}^x\phi(y)dy.$$
EO pricing under the model \eqref{model:CS} was investigated by
\cite{CS}. For a greater detail, see \cite{FanMancini2009}.  In this paper, we extend the
results of \cite{CS} by assuming a generalized SN distribution for the random variable $Z_\lambda$ in \eqref{model:CS} and call the  {\em generalized geometric skew Brownian motion}.\\

There are many extensions for the SN distribution. An
extension of the SN distribution was proposed by \cite{Azzalini1985} and
discussed  by \cite{ArnoldBeaver2002} in a more detail. More
specifically, the random variable $Z_{\lambda,\gamma}$ has a
generalized SN distribution with parameters
$\lambda,\gamma\in R$, denoted by $Z_{\lambda,\gamma}\sim
SN(\lambda,\gamma)$, if its pdf be
\begin{equation}\label{pdf:generalized:SN}
\phi(x;\lambda,\gamma)=\frac{\phi(x)\Phi(\lambda
x+\gamma)}{\Phi(\gamma/\sqrt{1+\lambda^2})},\ \ \
-\infty<x<+\infty,
\end{equation}
In the sequel sections, we assume that the stock price $\{S(t),\ t\geq 0 \}$
follows the generalized Azzalini skew Brownian motion, i.e.
\begin{equation}\label{behavior:st:under:P}
S(t)=S(0)\exp\left\{\mu
t+\sigma\sqrt{t}Z_{\lambda,\gamma}\right\},
\end{equation}
where $Z_{\lambda,\gamma}\sim SN(\lambda,\gamma)$. Notice that for $\gamma=0$, the random variable $Z_{\lambda,\gamma}$ with pdf \eqref{pdf:generalized:SN} is simplified to the Azzalini's skew normal and thus the model \eqref{behavior:st:under:P} is transformed to the model \eqref{model:CS}. Therefore, the results of this paper are extensions of the results provide by \cite{CS}. The rest of this article is organized as follow: In Section 2, the unique martingale measure is derived under the generalized geometric skew Brownian motion \eqref{behavior:st:under:P}. An explicit expression for the EO price is presented in Section 3. Some special cases are considered in more details in Section 4. In Section 5, the EO price's sensitivities to the skew parameters is considered. Section 6 concludes. The proofs are given in the appendix.
\section{Martingale measure}
The first step for obtaining the EO price is to
find an equivalent risk neutral probability measure, denoted by
$Q$, under which the discounted stock price process $\{e^{-rt}S(t)
\}$ is a martingale, where $r$ is the riskless continuous rate of
interest and $T$ is the expiry date of the option.\\

Let $M_X(a):=E(\exp\{a X\})$ denote the moment generating
function (MGF) of the random variable $X$. Under the model \eqref{behavior:st:under:P} with the objective
probability measure $P$, the MGF of the random variable
$\ln S(t)$ is derived as
\begin{eqnarray}
M_{\ln S(t)}(\beta)&=& E^{P}\left(e^{\beta\ln
S(t)}|\mathcal{F}_0\right)\nonumber\\
&=&\exp\{\beta\ln S(0)\} E^{P}\left(\exp\left\{\beta\left[\mu
t+\sigma\sqrt{t}Z_{\lambda,\gamma}\right]\right\}|\mathcal{F}_0\right)\nonumber\\
&=&\exp\{\beta\ln S(0)+\mu
t\} E^{P}\left(\exp\left\{\beta\sigma\sqrt{t}Z_{\lambda,\gamma}\right\}|\mathcal{F}_0\right)\nonumber\\
&=&\exp\{\beta\ln S(0)+\mu
t\}M_{Z_{\lambda,\gamma}}\left(\beta\sigma\sqrt{t}\right)\label{h1}
\end{eqnarray}
where $M_{Z_{\lambda,\gamma}}$ is the MGF of the random variable
$Z_{\lambda,\gamma}$ with pdf \eqref{pdf:generalized:SN} and $\mathcal{F}_0$ is the {\em information available to investors at present time $t=0$}. For more details, see \cite{KS}. \cite{ArnoldBeaver2002} showed that
\begin{equation}\label{MGF:generalized:SN}
M_{Z_{\lambda,\gamma}}(a)=\exp\left\{\frac{a^2}{2}\right\}\frac{\Phi\left(\frac{\gamma+\lambda
a }{\sqrt{1+\lambda^2}}\right)}{\Phi\left(\frac{\gamma
}{\sqrt{1+\lambda^2}}\right)}.
\end{equation}
From \eqref{h1} and \eqref{MGF:generalized:SN}, we have
\begin{eqnarray}
E^{P}\left(e^{-rt} S(t)|\mathcal{F}_0\right)&=& e^{-rt}
E^{P}\left( e^{\ln S(t)}|\mathcal{F}_0\right)\nonumber\\
&=& e^{-rt} M_{\ln S(t)}(1)\nonumber\\
&=& e^{-rt} \exp\{\ln S(0)+\mu
t\}M_{Z_{\lambda,\gamma}}\left(\sigma\sqrt{t}\right)\nonumber\\
&=& \exp\left\{\ln S(0)+\left(\mu-r+\frac{1}{2}\sigma^2\right)
t\right\}\frac{\Phi\left(\frac{\gamma+\lambda \sigma\sqrt{t}
}{\sqrt{1+\lambda^2}}\right)}{\Phi\left(\frac{\gamma
}{\sqrt{1+\lambda^2}}\right)}\nonumber\\
&=& S(0)\exp\left\{\left(\mu-r+\frac{1}{2}\sigma^2\right) t+\ln
\left[{\Phi\left(\frac{\gamma+\lambda \sigma\sqrt{t}
}{\sqrt{1+\lambda^2}}\right)}\bigg/{\Phi\left(\frac{\gamma
}{\sqrt{1+\lambda^2}}\right)}\right]\right\}.\nonumber\\ \label{h2}
\end{eqnarray}
From the identity $E^{Q}\left(e^{-rt}
S(t)|\mathcal{F}_0\right)=S(0)$ and \eqref{h2}, under the martingale measure $Q$, we have
\begin{equation}\label{behavior:st:under:Q}
S(t)=S(0)\exp\left\{\mu^\star
t+\sigma\sqrt{t}\tilde{Z}_{\lambda,\gamma}\right\},
\end{equation}
where
\[\mu^\star=r-\frac{1}{2}\sigma^2-\frac{1}{t}\ln
\left[{\Phi\left(\frac{\gamma+\lambda \sigma\sqrt{t}
}{\sqrt{1+\lambda^2}}\right)}\bigg/{\Phi\left(\frac{\gamma
}{\sqrt{1+\lambda^2}}\right)}\right],\] and
$\tilde{Z}_{\lambda,\gamma}$ has a generalized SN distribution
under the martingale probability measure $Q$. Notice that
\begin{eqnarray}
E^{Q}\left(e^{-rt} S(t)|\mathcal{F}_0\right)&=& e^{-rt}
E^{Q}\left( e^{\ln S(t)}|\mathcal{F}_0\right)\nonumber\\
&=& e^{-rt} M_{\ln S(t)}(1)\nonumber\\
&=& \exp\left\{\ln
S(0)+\left(\mu^\star-r+\frac{1}{2}\sigma^2\right)
t\right\}\frac{\Phi\left(\frac{\gamma+\lambda \sigma\sqrt{t}
}{\sqrt{1+\lambda^2}}\right)}{\Phi\left(\frac{\gamma
}{\sqrt{1+\lambda^2}}\right)}\nonumber\\
&=& S(0)\exp\left\{\left(\mu^\star-r+\frac{1}{2}\sigma^2\right)
t+\ln \left[{\Phi\left(\frac{\gamma+\lambda \sigma\sqrt{t}
}{\sqrt{1+\lambda^2}}\right)}\bigg/{\Phi\left(\frac{\gamma
}{\sqrt{1+\lambda^2}}\right)}\right]\right\}\nonumber\\
&=&S(0).
\end{eqnarray}
\section{European call option price}
Assuming a complete market, the equivalent martingale measure $Q$ will be unique (See, \cite{KS}). Therefore, the non-arbitrage (NA) EO price with the strike price $K$ and the expiration time $t$, denoted by $C(\mu,\sigma,\lambda,\gamma,r,t,K,S(0))$, is derived as
\begin{equation}\label{option:price:general}
C(\mu,\sigma,\lambda,\gamma,r,t,K,S(0))=e^{-rt}E^Q\left\{\left(S(t)-K\right)^{+}\bigg|\mathcal{F}_0\right\}.
\end{equation}
\begin{proposition}\label{proposition:EO:price:under:General:SN}
Let $S(t)=S(0)\exp\left\{\mu
t+\sigma\sqrt{t}Z_{\lambda,\gamma}\right\}$. Then
\begin{eqnarray}
C(\sigma,\lambda,\gamma,r,t,K,S(0))&=&S(0)\left\{1-\frac{\Phi_2\left(\frac{\lambda \sigma\sqrt{t}+\gamma
}{\sqrt{1+\lambda^2}},-w;\frac{-\lambda}{\sqrt{1+\lambda^2}}\right)}{\Phi\left(\frac{\lambda
\sigma\sqrt{t}+\gamma}{\sqrt{1+\lambda^2}}\right)}\right\}\nonumber\\
&&-e^{-rt} K\bar{\Phi}\left(-w+\sigma\sqrt{t};\lambda,\gamma\right),\label{option:price:general:skew}
\end{eqnarray}
where
\begin{equation}\label{w}
w=\frac{\log(S(0)/K)+(r+\frac{1}{2}\sigma^2) t-\ln
\left[{\Phi\left(\frac{\gamma+\lambda \sigma\sqrt{t}
}{\sqrt{1+\lambda^2}}\right)}\bigg/{\Phi\left(\frac{\gamma
}{\sqrt{1+\lambda^2}}\right)}\right] }{\sigma\sqrt{t}}.
\end{equation}
\end{proposition}
From Equations \eqref{option:price:general} and \eqref{option:price:general:skew}, we immediately see that
\begin{enumerate}
  \item $C$ is free of the drift parameter $\mu$.
  \item $C$ is an increasing, convex function of $S(0)$.
  \item $C$ is a decreasing, convex function of $K$.
  \item Since the value of an European call option is the same as that of an American call option (\cite{KS}), we conclude that $C$ is increasing in $t$.
\end{enumerate}
\begin{proposition}\label{proposition:EO:price:under:General:SN:limit:on:t}
$C\to S(0)$ as $t\to +\infty$.
\end{proposition}

\begin{corollary}
The non-arbitrage (NA) price of the European put option ($P$) is readily obtained from the well-known put-call parity formula (\cite{KS}, p.50), i.e.
\begin{equation}
P+C-S(0)=K e^{-rt},
\end{equation}
where $C$ is given by \eqref{option:price:general:skew}.
\end{corollary}

\section{Some special cases}
In what follows, we obtain simple expressions for
$C(\sigma,\lambda,\gamma,r,t,K,S(0))$ given by
\eqref{option:price:general:skew} in some special cases.
\subsection*{Case $\lambda=0$}
In this case, Equation \eqref{w} is reduced to
\begin{equation}\label{w:lambda:0}
w_1=\frac{\log(S(0)/K)+(r+\frac{1}{2}\sigma^2) t }{\sigma\sqrt{t}}.
\end{equation}
Also,
\begin{equation}\label{h5}
\frac{\Phi_2\left(\frac{\lambda \sigma\sqrt{t}+\gamma
}{\sqrt{1+\lambda^2}},-w_1;\frac{-\lambda}{\sqrt{1+\lambda^2}}\right)}{\Phi\left(\frac{\lambda
\sigma\sqrt{t}+\gamma}{\sqrt{1+\lambda^2}}\right)}=\frac{\Phi_2(\gamma,-w_1;0)}{\Phi(\gamma)}=\frac{\Phi(\gamma)\Phi(-w_1)}{\Phi(\gamma)}
=\Phi(-w_1).
\end{equation}
Upon substituting \eqref{w:lambda:0} and \eqref{h5} into \eqref{option:price:general:skew}, we have
\begin{equation}\label{h6}
C=S(0)\Phi(w_1)-e^{-r t}K\bar{\Phi}(-w_1+\sigma\sqrt{t};0,\gamma),
\end{equation}
where $w_1$ is given by \eqref{w:lambda:0}. From \eqref{pdf:generalized:SN}, one can see that $\Phi(x;0,\gamma)=\Phi(x)$ for every $\gamma\in(-\infty,+\infty)$. Thus, in this case we conclude from \eqref{h6} that
\begin{equation}\label{option:black:scholes}
C=S(0)\Phi(w_1)-e^{-r t}K\Phi(w_1-\sigma\sqrt{t}),
\end{equation}
which is the well known {\em Black-Scholes option price}, denoted by $C_{B-S}$.
\subsection*{Case $\gamma=0$}
In this case, we have from \eqref{w} that
\begin{equation}\label{w:gamma:0}
w_2=\frac{\log(S(0)/K)+(r+\frac{1}{2}\sigma^2) t-\ln
\left[2\Phi\left(\frac{\lambda \sigma\sqrt{t}
}{\sqrt{1+\lambda^2}}\right)\right] }{\sigma\sqrt{t}}.
\end{equation}
This implies from \eqref{option:price:general:skew} that
\begin{eqnarray}
C&=&S(0)\left\{1-\frac{\Phi_2\left(\frac{\lambda
\sigma\sqrt{t}
}{\sqrt{1+\lambda^2}},-w_2;\frac{-\lambda}{\sqrt{1+\lambda^2}}\right)}{\Phi\left(\frac{\lambda
\sigma\sqrt{t}}{\sqrt{1+\lambda^2}}\right)}\right\}-e^{-rt} K\bar{\Phi}\left(-w_2+\sigma\sqrt{t};\lambda,0\right),\nonumber\\
&=&S(0)\left\{1-\frac{\Phi_2\left(\frac{\lambda \sigma\sqrt{t}
}{\sqrt{1+\lambda^2}},-w_2;\frac{-\lambda}{\sqrt{1+\lambda^2}}\right)}{\Phi\left(\frac{\lambda
\sigma\sqrt{t}}{\sqrt{1+\lambda^2}}\right)}\right\}-e^{-rt}
K\bar{\Phi}_{SN}\left(-w_2+\sigma\sqrt{t};\lambda\right),\nonumber\\
\label{option:price:skew:gamma=0}
\end{eqnarray}
where $\Phi_{SN}(.;\lambda)$ stands for the cdf of the standard
Azzalini's skew normal distribution with parameter $\lambda$, i.e.
\[\Phi_{SN}(x;\lambda)=\int_{-\infty}^x 2\phi(y)\Phi(\lambda y)dy.\]
The option price \eqref{option:price:skew:gamma=0}, denoted by $C_{C-S}$, was obtained
by \cite{CS}. One may be noticed that they could
not find an explicit expression for
\eqref{option:price:skew:gamma=0} while we could presented EO price in \eqref{option:price:skew:gamma=0} in
terms of the cdf of the univariate skew normal and the bivariate
standard normal distributions. Numerical values of these
functions are provided by some statistical softwares such as $R$ with package $mnorm$.

%\subsection{Special case $\gamma=\lambda=0$}
%In this case, we have
%\begin{equation}\label{w:special:gamma=lambda=0}
%w_3=\frac{\log(S(0)/K)+(r+\frac{1}{2}\sigma^2) t}{\sigma\sqrt{t}}.
%\end{equation}
%Substituting \eqref{w:special:gamma=lambda=0} into
%\eqref{option:price:general:skew}, we have
%\begin{eqnarray}
%C&=&S(0)\left\{1-\frac{\Phi_2\left(0,-w_3;0\right)}{\Phi\left(0\right)}\right\}
%-e^{-rt} K\bar{\Phi}\left(-w_3+\sigma\sqrt{t};0,0\right),\nonumber
%\end{eqnarray}
%Since $\Phi(-x)=1-\Phi(x)$, we conclude
%\begin{eqnarray}
%C&=&S(0)\left\{1-\frac{\Phi\left(-w_3\right)/2}{1/2}\right\}
%-e^{-rt} K\bar{\Phi}\left(-w_3+\sigma\sqrt{t};0,0\right),\nonumber\\
%&=&S(0)\left\{1-\Phi\left(-w_3\right)\right\}
%-e^{-rt} K\Phi\left(w-\sigma\sqrt{t}\right),\nonumber\\
%&=&S(0)\Phi\left(w_3\right) -e^{-rt}
%K\Phi\left(w_3-\sigma\sqrt{t}\right),\nonumber
%\end{eqnarray}
%which is the well known {\em Black-Scholes option price formula}.

\section{Empirical evidences}
\begin{figure}
\centering
\includegraphics[width=1.5in,height=1.5in]{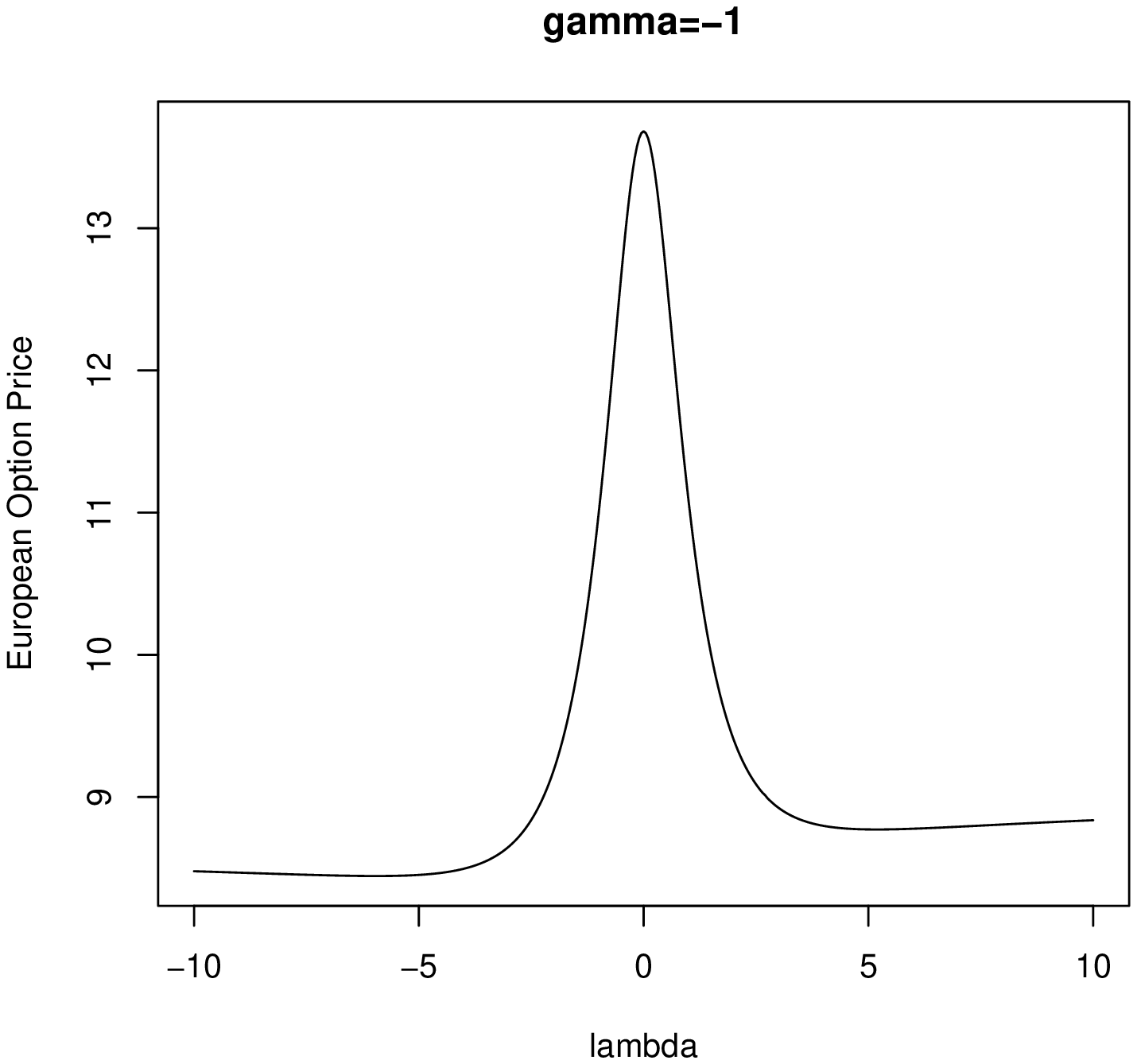}
\includegraphics[width=1.5in,height=1.5in]{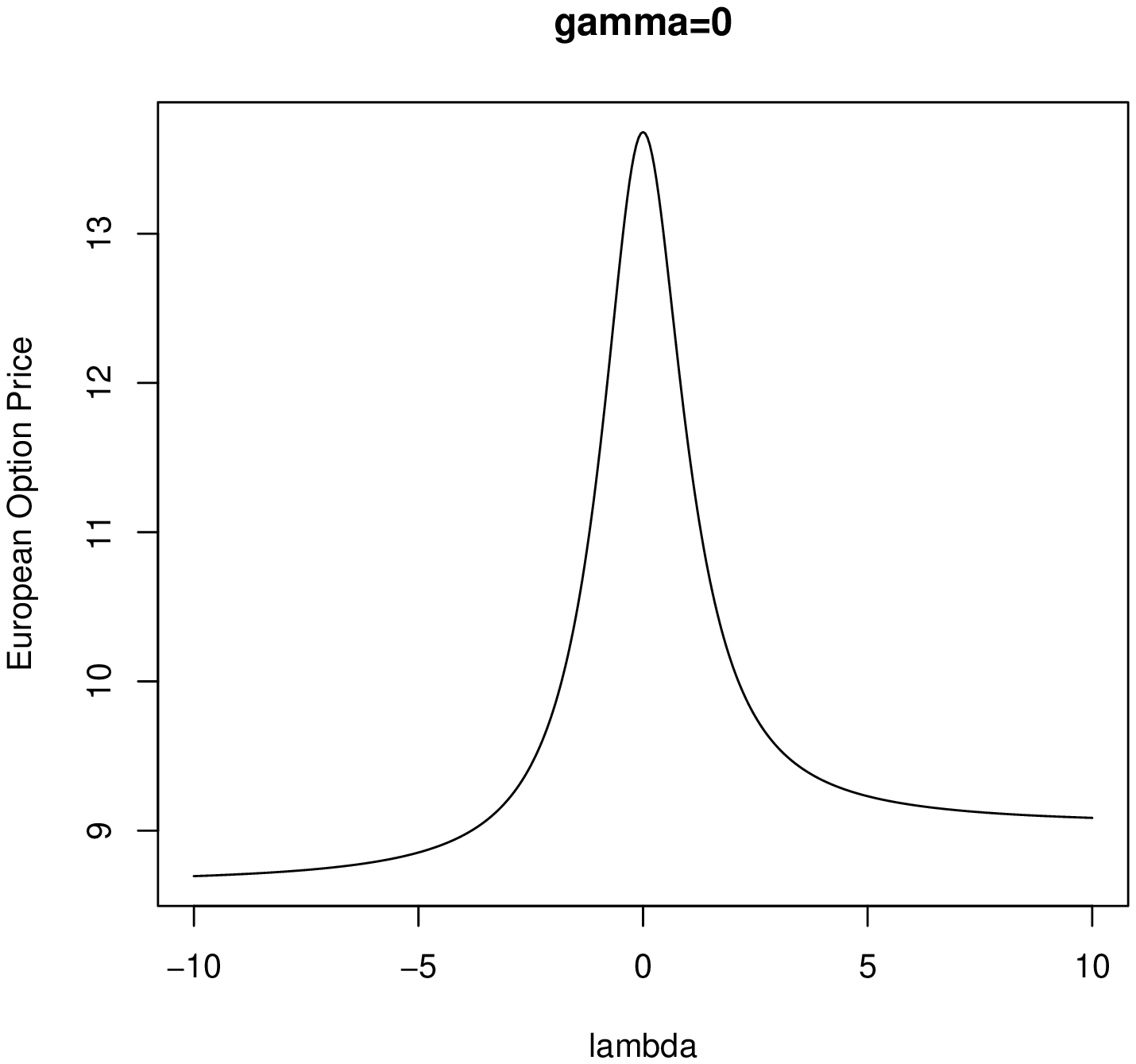}
\includegraphics[width=1.5in,height=1.5in]{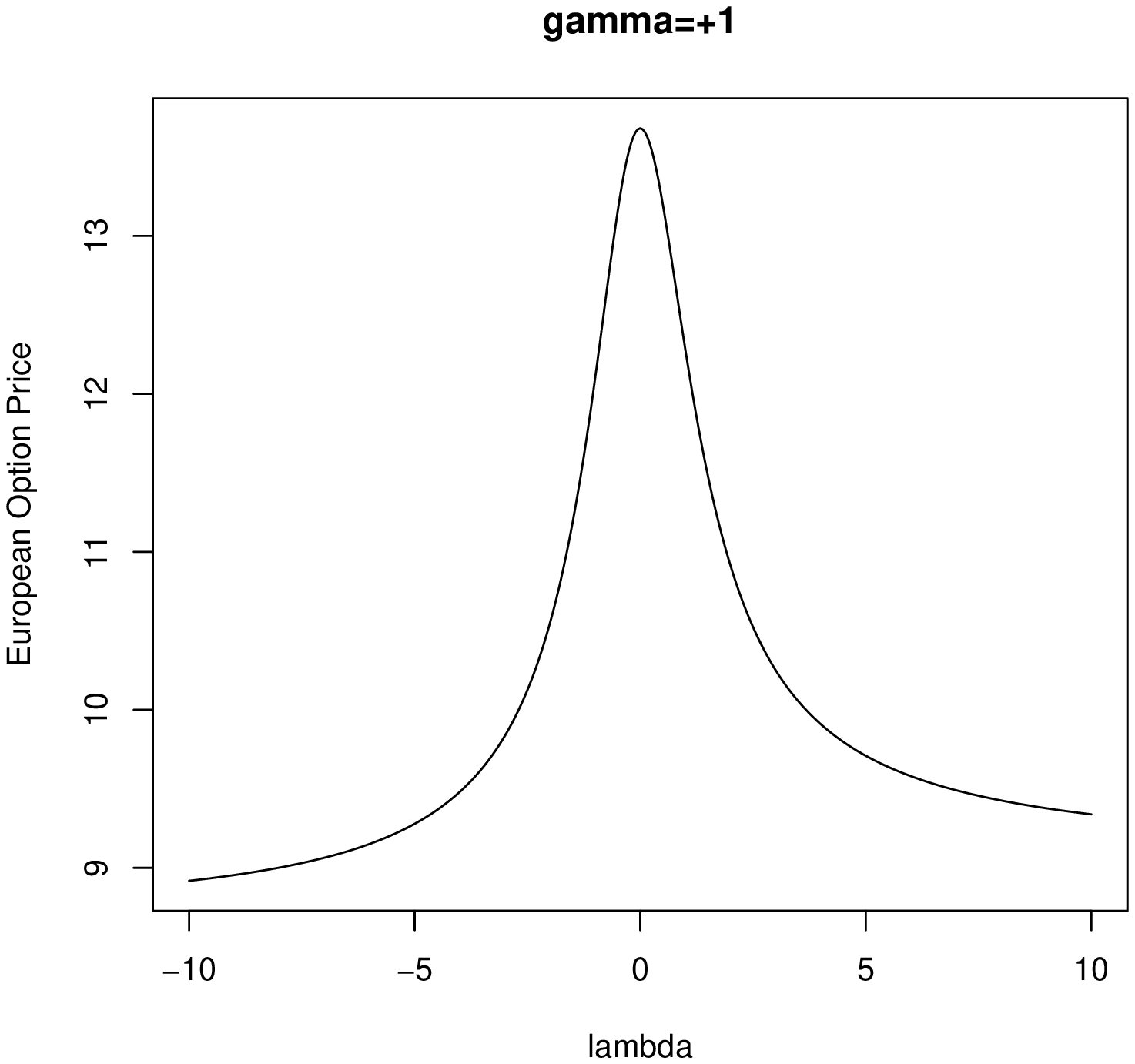}
\includegraphics[width=1.5in,height=1.5in]{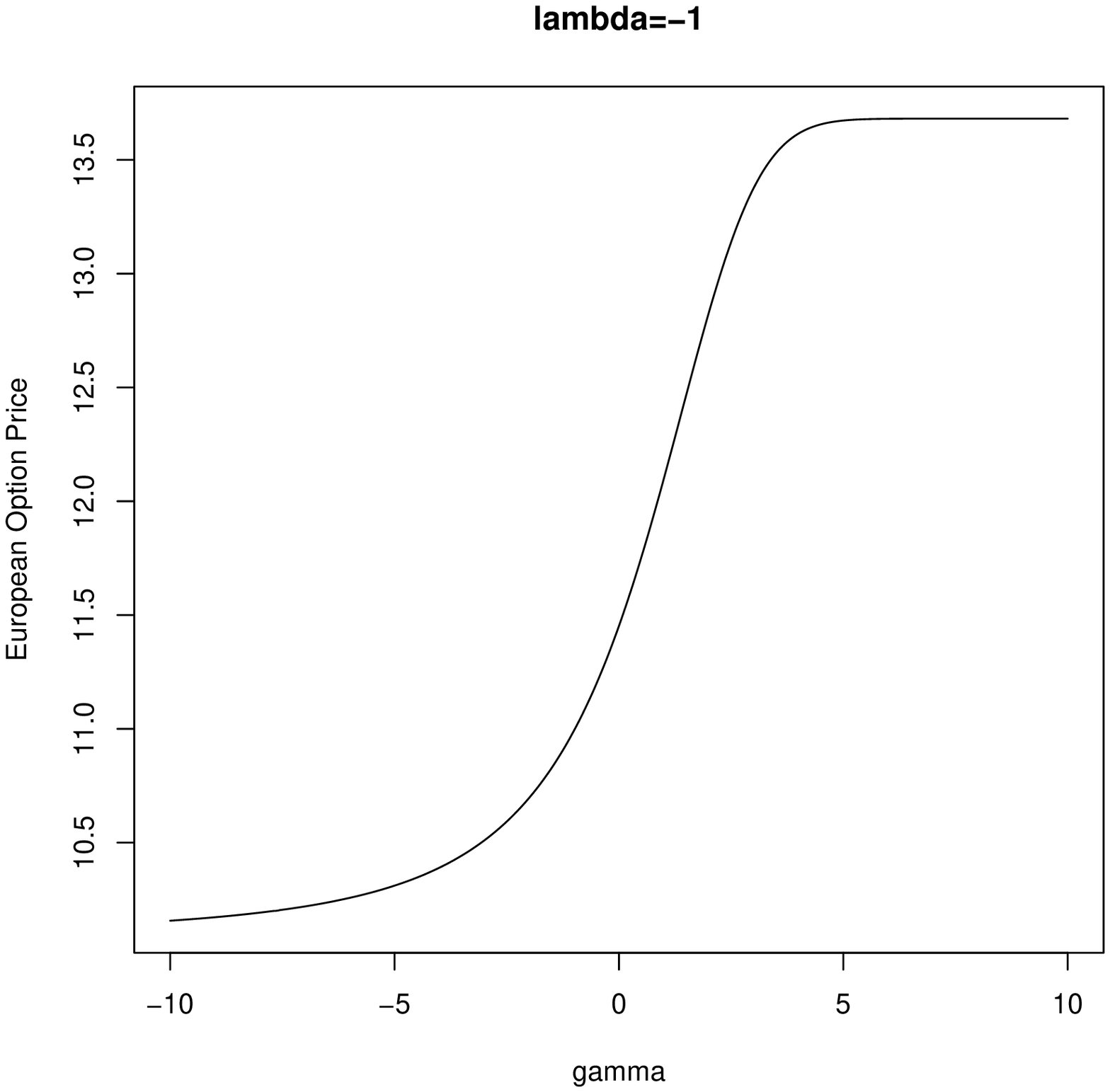}
\includegraphics[width=1.5in,height=1.5in]{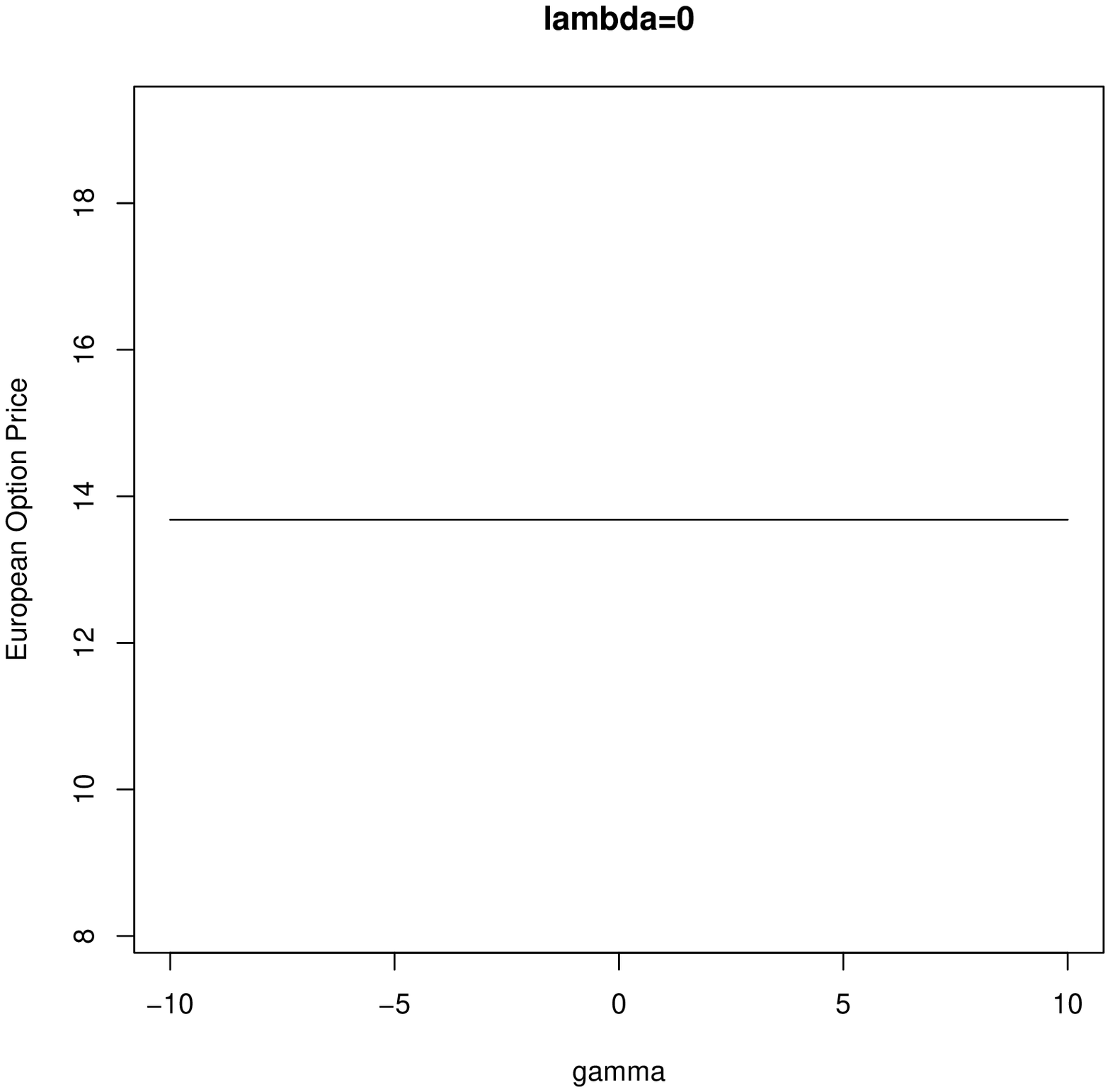}
\includegraphics[width=1.5in,height=1.5in]{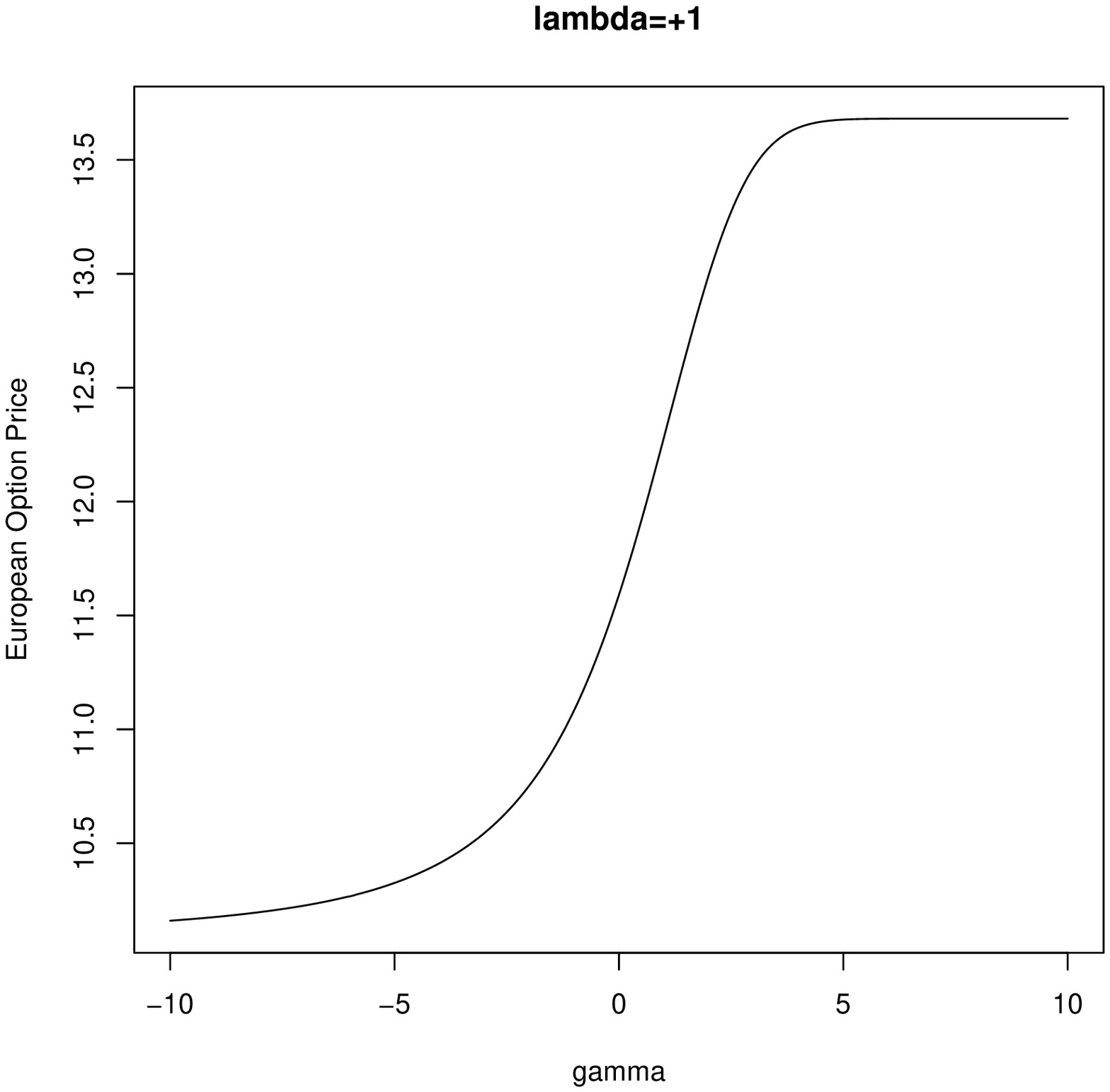}
\caption{Numerical values of C for the benchmark case and some selected values of the skew parameters $\lambda$ and $\gamma$.} \label{fig:model}
\end{figure}

In this section, we consider the option's sensitiveness to the skew parameters $\lambda$ and $\gamma$ in \eqref{option:price:general:skew} via a numerical methods because the partial derivatives of the option $C$ given by \eqref{option:price:general:skew} have complicated forms. The corresponding plots are given in Figure \ref{fig:model}. Following \cite{CS}, the benchmark case was taken as $S(0)=100$, $K=100$, $r=0.1$, $\sigma^2=0.4$ and $t=0.25$. Also, we consider $\lambda\in \{-1,0,1\}$ and $\gamma\in\{-1,0,1\}$.
Numerical values of $C$ for some selected values of the parameters are given in Table \ref{tab:C:some:values}.
\begin{table}
\caption{Numerical values of $C$ for some selected values of the skew parameters}
\centering
\begin{tabular}{cc|ccccc}\hline
        &&&&$\lambda$&&\\
        &   & -2&-1&0&+1&+2\\ \hline
        & -2& 8.702112 & 10.69672 & 13.68113        &10.75255&8.857459\\
        & -1& 9.188333 & 10.99278 & 13.68113        &11.08288&9.406439\\
$\gamma$&  0& 9.805336 & 11.45179 & {\bf 13.68113}  &11.59007&10.09846\\
        & +1& 10.55043 & 12.09882 & 13.68113        &12.27943&10.91346\\
        & +2& 11.37726 & 12.8264 &  13.68113        &12.99414&11.7723\\ \hline
\end{tabular}\label{tab:C:some:values}
\end{table}
\noindent Empirical evidences from Table \ref{tab:C:some:values} show that
\begin{itemize}
  \item the EO price $C$ is very sensitive w.r.t. the skew parameters $\lambda$ and $\gamma$;
  \item the EO price $C$ is decreasing in $|\lambda|$;
  \item for $\lambda=0$, Black-Scholes EO price $C_{B-S}$ in Equation \eqref{option:black:scholes} is obtained;
  \item for $\lambda\neq 0$, we have an over estimation by Black-Scholes pricing EO $(C<C_{BS})$, that is, overpricing by Black-Scholes EO pricing leads to out-of-money;
   \item for $\lambda=0$, the option price $C$ does not depend on the parameter $\gamma$ while for $\lambda\neq 0$, the option price $C$ is increasing in the parameter $\gamma$;
  \item for $\lambda\neq 0$ and $\gamma>(<)0$ we have $C>(<)C_{C-S}$, that is $C_{C-S}$ leads to in-the-money (out-of-money).
\end{itemize}
\section{Conclusions}
Assuming a generalized SN model, the problem of European option pricing was investigated. Existence and uniqueness of the martingale measure was shown. The explicit expression for EO price was derived in terms of the cdfs of the univariate SN and the bivariate standard normal distributions. Empirically, it was shown that the EO price is sensitive to the skew parameters. The results obtained may be extended for other general SN models. See for example \cite{ShafieiDoostBala2012}. Another important topic is the problem of the estimating skew parameters on the basis of the observed EO prices \cite{CorradoSu1997}. Work in this direction is currently under progress and we hope to report findings in a future paper.

\section*{Appendix}
\subsection*{Proof of Proposition \ref{proposition:EO:price:under:General:SN}}

From  \eqref{behavior:st:under:Q} and \eqref{option:price:general}, we have
\begin{eqnarray*}
e^{rt}C(\mu,\sigma,\lambda,\gamma,r,t,K)&=&E^Q\left\{\left(S(t)-K\right)^{+}\bigg|\mathcal{F}_0\right\}\\
&=&E^Q\left\{\left(S(0)\exp\left\{\mu^\star t+\sigma\sqrt{t}\tilde{Z}_{\lambda,\gamma}\right\}-K\right)^{+}\bigg|\mathcal{F}_0\right\}\\
&=&\int_{-\infty}^{+\infty}\left(S(0)\exp\left\{\mu^\star
t+\sigma\sqrt{t}x\right\}-K\right)^{+}\phi(x;\lambda,\gamma) dx\\
&=&\int_{S(0)\exp\left\{\mu^\star
t+\sigma\sqrt{t}x\right\}>K}\left[S(0)\exp\left\{\mu^\star
t+\sigma\sqrt{t}x\right\}-K\right]\phi(x;\lambda,\gamma)
dx\\
&=&\int_{\frac{\log(K/S(0))-\mu^\star t
}{\sigma\sqrt{t}}}^{+\infty}\left[S(0)\exp\left\{\mu^\star
t+\sigma\sqrt{t}x\right\}-K\right]\phi(x;\lambda,\gamma)
dx\\
&=&S(0)\exp\left\{\mu^\star t\right\}\int_{\frac{\log(K/S(0))-\mu^\star t
}{\sigma\sqrt{t}}}^{+\infty}\exp\left\{\sigma\sqrt{t}x\right\}\phi(x;\lambda,\gamma)
dx\\
&&-K\bar{\Phi}\left(\frac{\log(K/S(0))-\mu^\star t
}{\sigma\sqrt{t}};\lambda,\gamma\right),
\end{eqnarray*}
where $\bar{\Phi}(x;\lambda,\gamma)=P(Z_{\lambda,\gamma}>x)$.
Thus,
\begin{equation}\label{option:price:1}
e^{rt}C(\mu,\sigma,\lambda,\gamma,r,t,K)=S(0)\exp\left\{\mu^\star
t\right\}B-K\bar{\Phi}\left(a;\lambda,\gamma\right),
\end{equation}
where $\displaystyle B=\int_{-\infty}^{+\infty}
I_{[a,+\infty]}(x)\exp\left\{\sigma\sqrt{t}x\right\}\phi(x;\lambda,\gamma)
dx$ and $a=\frac{\log(K/S(0))-\mu^\star t }{\sigma\sqrt{t}}$ and
$I_A(x)$ is the indicator function for the set $A$, i.e.
\[I_A(x)=\left\{
\begin{array}{ccc}
1,&\ \ & x\in A\\
0,&& o.w.
\end{array}
\right.
\]
Therefore,
\begin{eqnarray}
B&=&E^Q\left[I_{[a,+\infty]}(Z_{\lambda,\gamma})\exp\left\{\sigma\sqrt{t}Z_{\lambda,\gamma}\right\}\right]\nonumber\\
&=&\left[1-\Phi(a;\lambda,\gamma)\right]E^Q\left[I_{[a,+\infty]}(Z_{\lambda,\gamma})
\frac{\exp\left\{\sigma\sqrt{t}Z_{\lambda,\gamma}\right\}}
{1-\Phi(a;\lambda,\gamma)}\right]\nonumber\\
&=&\left[1-\Phi(a;\lambda,\gamma)\right]E^Q\left[\exp\left\{\sigma\sqrt{t}Z_{T(a,+\infty);\lambda,\gamma}\right\}\right]\nonumber\\
&=&\left[1-\Phi(a;\lambda,\gamma)\right]MGF(\sigma\sqrt{t};a,+\infty,\lambda,\gamma),\label{B:expression:1}
\end{eqnarray}
where $Z_{T(a,b);\lambda,\gamma}$ is the truncated skew normal
distribution to interval $[a,b]$, introduced by Jamalizadeh {\em
et al.}(2009), and $MGF(s;a,b,\lambda,\gamma)$ is the
corresponding MGF. Jamalizadeh {\em et
al.}(2009) derived an explicit expression for
$MGF(s;a,b,\lambda,\gamma)$ as
\begin{eqnarray}
MGF(s;a,b,\lambda,\gamma)&=&u(\lambda,\gamma,a,b)e^{s^2/2}\left\{\Phi_2\left(\frac{\lambda
s+\gamma
}{\sqrt{1+\lambda^2}},b-s;\frac{-\lambda}{\sqrt{1+\lambda^2}}\right)\right.\nonumber\\
&&-\left.\Phi_2\left(\frac{\lambda s+\gamma
}{\sqrt{1+\lambda^2}},a-s;\frac{-\lambda}{\sqrt{1+\lambda^2}}\right)\right\},\label{mgf:TSN}
\end{eqnarray}
where
\begin{equation}\label{mgf:TSN:constant}
\left[u(\lambda,\gamma,a,b)\right]^{-1}=\Phi\left(\frac{\gamma}{\sqrt{1+\lambda^2}}\right)\left\{
\Phi_{SN}(b;\lambda,\gamma)-\Phi_{SN}(a;\lambda,\gamma)\right\},
\end{equation} and $\Phi_2(.,.;\delta)$ is the cdf of
$N_2(0,0,1,1,\delta)$ (the standard bivariate normal distribution
with correlation coefficient $\delta$). From
\eqref{B:expression:1}, \eqref{mgf:TSN} and
\eqref{mgf:TSN:constant}, one can easily show that
\begin{eqnarray}
B&=&\bar{\Phi}(a;\lambda,\gamma)MGF(\sigma\sqrt{t};a,+\infty,\lambda,\gamma)\nonumber\\
&=&\bar{\Phi}(a;\lambda,\gamma)u(\lambda,\gamma,a,+\infty)e^{(\sigma\sqrt{t})^2/2}\left\{\Phi_2\left(\frac{\lambda
\sigma\sqrt{t}+\gamma
}{\sqrt{1+\lambda^2}},+\infty;\frac{-\lambda}{\sqrt{1+\lambda^2}}\right)\right.\nonumber\\
&&-\left.\Phi_2\left(\frac{\lambda \sigma\sqrt{t}+\gamma
}{\sqrt{1+\lambda^2}},a-\sigma\sqrt{t};\frac{-\lambda}{\sqrt{1+\lambda^2}}\right)\right\}\nonumber\\
&=&\frac{1}{\Phi\left(\frac{\gamma}{\sqrt{1+\lambda^2}}\right)}e^{\sigma^2
t/2}\left\{\Phi\left(\frac{\lambda \sigma\sqrt{t}+\gamma
}{\sqrt{1+\lambda^2}}\right)-\Phi_2\left(\frac{\lambda
\sigma\sqrt{t}+\gamma
}{\sqrt{1+\lambda^2}},a-\sigma\sqrt{t};\frac{-\lambda}{\sqrt{1+\lambda^2}}\right)\right\}.\nonumber\\
&&\label{B:expression:2}
\end{eqnarray}
Upon substituting \eqref{B:expression:2} into
\eqref{option:price:1},
we have
\begin{eqnarray}
C(\mu,\sigma,\lambda,\gamma,r,t,K)&=&e^{-rt}\left[\frac{S(0)e^{\mu^\star
t+ \sigma^2
t/2}}{\Phi\left(\frac{\gamma}{\sqrt{1+\lambda^2}}\right)}\left\{\Phi\left(\frac{\lambda
\sigma\sqrt{t}+\gamma
}{\sqrt{1+\lambda^2}}\right)\right.\right.\nonumber\\
&&\left.\left.-\Phi_2\left(\frac{\lambda \sigma\sqrt{t}+\gamma
}{\sqrt{1+\lambda^2}},\frac{\log(K/S(0))-\mu^\star t }{\sigma\sqrt{t}}-\sigma\sqrt{t};\frac{-\lambda}{\sqrt{1+\lambda^2}}\right)\right\}\right.\nonumber\\
&&\left.-K\bar{\Phi}\left(\frac{\log(K/S(0))-\mu^\star t
}{\sigma\sqrt{t}};\lambda,\gamma\right)\right]\nonumber\\
&=&e^{-rt}\left[\frac{S(0)e^{rt}}{\Phi\left(\frac{\lambda
\sigma\sqrt{t}+\gamma}{\sqrt{1+\lambda^2}}\right)}\left\{\Phi\left(\frac{\lambda
\sigma\sqrt{t}+\gamma
}{\sqrt{1+\lambda^2}}\right)\right.\right.\nonumber\\
&&\left.\left.-\Phi_2\left(\frac{\lambda \sigma\sqrt{t}+\gamma
}{\sqrt{1+\lambda^2}},\frac{\log(K/S(0))-\mu^\star t }{\sigma\sqrt{t}}-\sigma\sqrt{t};\frac{-\lambda}{\sqrt{1+\lambda^2}}\right)\right\}\right.\nonumber\\
&&\left.-K\bar{\Phi}\left(\frac{\log(K/S(0))-\mu^\star t
}{\sigma\sqrt{t}};\lambda,\gamma\right)\right]\nonumber\\
&=&S(0)\left\{1-\frac{\Phi_2\left(\frac{\lambda \sigma\sqrt{t}+\gamma
}{\sqrt{1+\lambda^2}},\frac{\log(K/S(0))-\mu^\star t }{\sigma\sqrt{t}}-\sigma\sqrt{t};\frac{-\lambda}{\sqrt{1+\lambda^2}}\right)}{\Phi\left(\frac{\lambda
\sigma\sqrt{t}+\gamma}{\sqrt{1+\lambda^2}}\right)}\right\}\nonumber\\
&&-e^{-rt} K\bar{\Phi}\left(-w+\sigma\sqrt{t};\lambda,\gamma\right)\label{h4}
\end{eqnarray}
where
\[w=-\frac{\log(K/S(0))-\mu^\star t }{\sigma\sqrt{t}}+\sigma\sqrt{t}.\]
After some algebraic manipulations, a simplified version for $w$
is derived as
\begin{eqnarray}
w&=&-\frac{\log(K/S(0))-\mu^\star t }{\sigma\sqrt{t}}+\sigma\sqrt{t}\nonumber\\
&=&\frac{\log(S(0)/K)+(r-\frac{1}{2}\sigma^2) t-\ln
\left[{\Phi\left(\frac{\gamma+\lambda \sigma\sqrt{t}
}{\sqrt{1+\lambda^2}}\right)}\bigg/{\Phi\left(\frac{\gamma
}{\sqrt{1+\lambda^2}}\right)}\right] }{\sigma\sqrt{t}}+\sigma\sqrt{t}\nonumber\\
&=&\frac{\log(S(0)/K)+(r+\frac{1}{2}\sigma^2) t-\ln
\left[{\Phi\left(\frac{\gamma+\lambda \sigma\sqrt{t}
}{\sqrt{1+\lambda^2}}\right)}\bigg/{\Phi\left(\frac{\gamma
}{\sqrt{1+\lambda^2}}\right)}\right] }{\sigma\sqrt{t}},\label{w:first:expression}
\end{eqnarray}
and the desired result follows from \eqref{h4} and \eqref{w:first:expression}.\hfill{$\Box$}
\subsection*{Proof of Proposition \ref{proposition:EO:price:under:General:SN:limit:on:t}}
From \eqref{w}, we have $\displaystyle \lim_{t\to+\infty}w=+\infty$ and
\[\lim_{t\to+\infty}e^{-rt} K\bar{\Phi}\left(-w+\sigma\sqrt{t};\lambda,\gamma\right)=0.\]
Also,
\[\lim_{t\to+\infty}\frac{\Phi_2\left(\frac{\lambda \sigma\sqrt{t}+\gamma
}{\sqrt{1+\lambda^2}},-w;\frac{-\lambda}{\sqrt{1+\lambda^2}}\right)}{\Phi\left(\frac{\lambda
\sigma\sqrt{t}+\gamma}{\sqrt{1+\lambda^2}}\right)},\]
and the desired result follows from \eqref{option:price:general:skew}. \hfill{$\Box$}

\end{document}